\documentclass{article}
\usepackage{graphicx}
\usepackage{amsmath}
\usepackage{amsfonts}

\title{A non-associative quantum mechanics}
\author{\textbf{Vladimir Dzhunushaliev}}
\date{}

\begin{document}
\maketitle

\begin{center}
\textit{
Dept. Phys. and Microel. Engineer., Kyrgyz-Russian
Slavic University, Bishkek, Kievskaya Str. 44, 720021, Kyrgyz
Republic \\ 
e-mail: dzhun@hotmail.kg}
\end{center}

%\pacs{}
\vspace{24pt}
\begin{abstract}
A non-associative quantum mechanics is proposed in which the product of three and
more operators can be non-associative one. The multiplication rules of the
octonions define the multiplication rules of the corresponding operators with quantum
corrections. The self-consistency of the operator algebra is proved for the product
of three operators. Some properties of the non-associative quantum
mechanics are considered. It is proposed that some generalization of the
non-associative algebra of quantum operators can be helpful for understanding
of the algebra of field operators with a strong interaction.
\end{abstract}
\vspace{24pt}

%\pacs{12.38.Aw, 12.38.Lg}

\maketitle

\vspace{24pt}
\section{INTRODUCTION}
\vspace{12pt}

In any modern quantum theory the algebra of quantized fields is defined by canonical
commutations/anticommutations relationships between field operators in two
different points. It is well known that strictly speaking such algebra can correctly
describe the algebra of free fields only. For weakly interacting fields (quantum electrodynamics, for example) we can calculate the n-points Green's
functions using Feynman diagram technique and in principle these functions
give us full information about interacting fields. But for the strongly
interacting fields the situation is another: without the
knowledge of the algebra of quantized fields we can not carried out the operator
quantization of such theories. The examples of strongly interacting theories are: 
$\lambda \phi^4$-- theory with $\lambda \gg 1$, quantum chromodynamics in some regimes and gravity at the Planck level. Therefore one can formulate one of the main
problems in quantum field theory: the formulation of the algebra of quantized
fields whose generating relationships are more complicated the canonical
relationships. Of course such problem
for quantum field theory is very complicated and in this notice we would
like to present a non-associative quantum mechanics where we have two
relationships generating the algebra of quantum operators: the first one
is similar to the ordinary anticommutation rule for the creation and annihilation
fermion operators and the second one is the rule for the movement of brackets
in the product of three operators.

\vspace{24pt}
\section{NON-ASSOCIATIVITY: PRELIMINARY \\ REMARKS}
\vspace{12pt}

In this section we would like to remember for reader the first example of a
non-associativity in mathematics: octonions and review some attempts to
introduce the non-associativity in physics.
\par
At first we would like to give some preliminary in the non-associativity.
A vector space is a finite-dimensional module over the
field of real numbers.  An algebra $A$ is a vector space which
is equipped with a bilinear map $m : A \times A \to A$
(multiplication) and a nonzero element $1 \in A$ (unit) such
that $1 \cdot a = a \cdot 1 = a$.  The algebra $A$ is non-associative one if
\begin{equation}
    \left( a \cdot b \right) \cdot c \neq a \cdot \left( b \cdot c \right).
\label{sec1-10}
\end{equation}
The algebra $A$ is alternative if
\begin{equation}
    \left( a \cdot a \right) \cdot b = a \cdot \left( a \cdot b \right),
    \quad
    \left( b \cdot a \right) \cdot a = b \cdot \left( a \cdot a \right),
    \quad
    \left( a \cdot b \right) \cdot a = a \cdot \left( b \cdot a \right).
\label{sec1-20}
\end{equation}
Later we will omit the multiplication sign $(\cdot)$.

\subsection{The Octonions}
\vspace{12pt}

Good and simple introduction for physicists to the octonions is Ref.\cite{baez}. The octonions are an 8D algebra with the basis
$1, \tilde{q}_i, \tilde{Q}_i, I$ and their multiplication is given in the
Table \ref{oct}.
\begin{table}[h]
\begin{tabular}{|c|c|c|c|c|c|c|c|c|}                                                \hline
&$\tilde{q}_1$ & $\tilde{q}_2$ & $\tilde{q}_3$ &
$\tilde{Q}_1$  & $\tilde{Q}_2$ & $\tilde{Q}_3$ & $I$         \\ \hline
$\tilde{q}_1$ & $-1  $ & $-\tilde{q}_3$  & $\tilde{q}_2$  &
                $-I$ & $-\tilde{Q}_3$  & $\tilde{Q}_2$ & $\tilde{Q}_1$ \\ \hline
$\tilde{q}_2$ & $\tilde{q}_3$ & $-1$   & $-\tilde{q}_1$  &
                        $\tilde{Q}_3$  & $-I$ & $-\tilde{Q}_1$  & $\tilde{Q}_2$    \\ \hline
$\tilde{q}_3$ & $-\tilde{q}_2$ & $\tilde{q}_1$ & $-1$   &
                        $-\tilde{Q}_2$  & $\tilde{Q}_1$  & $-I$ & $\tilde{Q}_3$    \\ \hline
$\tilde{Q}_1$ & $I $ & $-\tilde{Q}_3$ & $\tilde{Q}_2$ &
                        $ 1$   & $\tilde{q}_3$  & $-\tilde{q}_2$  & $\tilde{q}_1$    \\ \hline
$\tilde{Q}_2$ & $\tilde{Q}_3$ & $I$  & $-\tilde{Q}_1$ &
                        $-\tilde{q}_3$ & $ 1$   & $\tilde{q}_1$  & $\tilde{q}_2$    \\ \hline
$\tilde{Q}_3$ & $-\tilde{Q}_2 $ & $\tilde{Q}_1$ & $I$  &
                        $\tilde{q}_2$ & $-\tilde{q}_1$ & $ 1$   & $\tilde{q}_3$    \\ \hline
                    $I$ & $-\tilde{Q}_1 $ & $-\tilde{Q}_2$  & $-\tilde{Q}_3$ &
                    $-\tilde{q}_1$  & $-\tilde{q}_2$ & $-\tilde{q}_3$ & $ 1$    \\ \hline
\end{tabular}
\caption{Octonion multiplication table}
\label{oct}
\end{table}
It is easy to see that this algebra is anticommutative
\begin{equation}
    ab = - ba, \quad a,b = \tilde{q}_i, \tilde{Q}_i, I,
    \quad i = 1,2,3
\label{sec1-30}
\end{equation}
and sometimes non-associative, for example
\begin{eqnarray}
    &&\ \ \tilde{Q}_m \left( \tilde{Q}_n \tilde{Q}_p \right) =
    - \left( \tilde{Q}_m \tilde{Q}_n  \right) \tilde{Q}_p ,
    \quad m \neq n, n \neq p, p \neq m,
\label{sec1-40}\\
    &&\left.
    \begin{array}{rcl}
        \tilde{q}_m \left( \tilde{Q}_n \tilde{q}_n \right) &=&
                - \left( \tilde{q}_m \tilde{Q}_n  \right) \tilde{q}_n , \\
        \tilde{q}_m \left( \tilde{Q}_m \tilde{q}_n \right) &=&
                - \left( \tilde{q}_m \tilde{Q}_m  \right) \tilde{q}_n , \\
            \tilde{Q}_m \left( \tilde{q}_n \tilde{Q}_n \right) &=&
                - \left( \tilde{Q}_m \tilde{q}_n  \right) \tilde{Q}_n ,\\
            \tilde{Q}_m \left( \tilde{q}_m \tilde{Q}_n \right) &=&
                - \left( \tilde{Q}_m \tilde{q}_m  \right) \tilde{Q}_n  \\
    \end{array}
    \right\}
    \; m \neq n.
\label{sec1-50}
\end{eqnarray}
The octonions designations in this paper $\tilde{q}_i, \tilde{Q}_i, i=1,2,3$ are coordinated with the octonions designations $e_i, i=1, \ldots , 7$ in Ref.\cite{baez} as follows: 
\begin{equation}
  \tilde{q}_1 = e_1, \tilde{q}_2 = e_3, \tilde{q}_3 = e_7, 
  \tilde{Q}_1 = ie_5, \tilde{Q}_2 = ie_4, \tilde{Q}_3 = ie_2 
  \quad \text{here  } i = \sqrt{-1}
\label{sec1-52}
\end{equation}
and with the designations in Ref.\cite{merab} as follows: 
\begin{equation}
  \tilde{q}_i = j_i, \tilde{Q}_i = J_i, i=1,2,3.
\label{sec1-54}
\end{equation}

\subsection{Non-associativity in Physics}
\vspace{12pt}

Probably the first attempt to apply the octonions in physics was
the paper \cite{jordan} describing an octonionic quantum
mechanics. In Ref.\cite{kugo} it is shown that some relation
between octonions and string theory exists. An octonionic geometry
in Ref.\cite{merab} is described. In this paper a new geometrical
interpretation of the products of octonionic basis units is
presented and eight real parameters of octonions are interpreted
as the spacetime coordinates, momentum and energy. In Ref.\cite{dzhun} the non-associativity for the non-perturbative
quantization of strongly interacting fields is introduced. A
generalization of the quantum Hall effect where particles move in
an eight dimensional space under an $SO(8)$ gauge field in Ref.\cite{Bernevig:2003yz} is considered. The underlying mathematics
of this particle liquid is that of the last normed division
algebra, the octonions. In Ref.\cite{Nesterov:2000qb} in the
framework of non-associative geometry a unified description of
continuum and discrete spacetime is proposed. In Ref(s).\cite{Grossman},
\cite{Jackiw} it is shown that for the generators of translations
the Jacobi identity in the presence of the monopole fails and for
the finite translations the operators $U(\vec{a}) \equiv
\exp\left( i \vec{a} \hat{\vec{p}} \right)$ do not associate. In
Ref.\cite{Nesterov:2004bn} it is shown that non-associative
extension of the group U(1) allows to obtain a consistent theory
of point-like magnetic monopole with an arbitrary magnetic charge.
In Ref.\cite{Color} a possible connection between quark
confinement and octonions are investigated.

\vspace{24pt}
\section{A NON-ASSOCIATIVE QUANTUM \\ MECHANICS}
\vspace{12pt}

In this section we would like to show that quantum mechanics can be based
not only on the non-commutativity of operators but on the non-associativity, too.
The main aim for such construction is that it can be a first step to building
a non-perturbative quantum theory where the algebra of field operators is
more complicated the algebra formed with the canonical non-commutative relationships.
\par
Our basic idea is that the classical definition of an associator of three numbers
$a,b,c$
\begin{equation}
  (ab)c - a(bc) = Ass\left[ a(bc) \right]
\label{sec2-10}
\end{equation}
which measures the failure of associativity, can be generalized for operators
\begin{equation}
  \left(\hat{a} \hat{b}\right)\hat{c} - \hat{a} \left(\hat{b} \hat{c}\right) =
  Ass\left[ \hat{a} \left(\hat{b} \hat{c}\right) \right] .
\label{sec2-20}
\end{equation}
where the quantity $Ass\left[ \hat{a} \left(\hat{b} \hat{c}\right) \right]$
is received from $Ass\left[ a(bc) \right]$ at the replacement
$a,b,c \rightarrow \hat{a}, \hat{b}, \hat{c}$ and  quantum corrections are
introduced. Of course the associator $Ass \left[ \hat{a} (\hat{b} \hat{c}) \right]$
depends on the product $\hat{a} \left(\hat{b} \hat{c}\right)$.
\par
For example, the standard quantization procedure (on the level of creation/annihilaion operators) can be presented as the destruction
of the multiplication rule of real
\begin{equation}
    a b = b a
\label{sec2-50}
\end{equation}
or Grassman numbers
\begin{equation}
    a b = - b a
\label{sec2-60}
\end{equation}
by following wayt:
\begin{equation}
  \hat{a} \hat{b} \mp \hat{b} \hat{a}= 1 .
\label{sec2-70}
\end{equation}
Following by this prescription we have to destroy the multiplication rules
\eqref{sec1-30} for the product of two octonions and \eqref{sec1-40}
\eqref{sec1-50} for the product of three octonions.
For brevity we will omit the sign $(\:\hat{}\:)$. For the product of two
operators with $i \neq j $ we will have
\begin{eqnarray}
    q_i q_j &=& - q_j q_i,
\label{sec2-72}\\
    q_i Q_j &=& - Q_j q_i,
\label{sec2-74}\\
    Q_i Q_j &=& - Q_j Q_i
\label{sec2-80}
\end{eqnarray}
and for the identical indexes
\begin{equation}
  q_i Q_i + Q_i q_i = 1
\label{sec2-90}
\end{equation}
which is the ordinary canonical anti-commutation relationships for the  creation/annihilation fermion operators and other anti-commutators are zero similar to ordinary quantum field theory. The associators/anti-associators for the product of three operators with different indexes
$m \neq n, n \neq p, p \neq m $ are
\begin{eqnarray}
    Q_m\left( Q_n Q_p \right) &=& - \left( Q_mQ_n \right) Q_p +
    \epsilon_{mnp} \mathcal{H}_3 ,
\label{sec2-100}\\
    q_m\left( q_n q_p \right) &=& \left( q_m q_n \right) q_p +
    \epsilon_{mnp} \mathcal{H}_3 ,
\label{sec2-110}\\
    q_m\left( Q_n q_p \right) &=& -\left( q_m Q_n \right) q_p +
    \epsilon_{mnp} \mathcal{H}_3 ,
\label{sec2-120}\\
    q_m\left( q_n Q_p \right) &=& \left( q_m q_n \right) Q_p +
    \epsilon_{mnp} \mathcal{H}_3 ,
\label{sec2-130}\\
    Q_m\left( q_n q_p \right) &=& \left( Q_m q_n \right) q_p +
    \epsilon_{mnp} \mathcal{H}_3 ,
\label{sec2-140}\\
    q_m\left( Q_n Q_p \right) &=& \left( q_m Q_n \right) Q_p +
    \epsilon_{mnp} \mathcal{H}_3 ,
\label{sec2-150}\\
    Q_m\left( q_n Q_p \right) &=& \left( Q_m q_n \right) Q_p +
    \epsilon_{mnp} \mathcal{H}_3 ,
\label{sec2-160}\\
    Q_m\left( Q_n q_p \right) &=& \left( Q_m Q_n \right) q_p +
    \epsilon_{mnp} \mathcal{H}_3
\label{sec2-170}
\end{eqnarray}
where $\mathcal{H}_3$ is for the time undefined operator, the index $3$ of the operator
$\mathcal{H}_3$ means that we consider the product of three operators. The associators/anti-associators for the product of three operators as $q(Qq)$ and $Q(qQ)$ and with two different indexes $m \neq n$ we have
\begin{eqnarray}
    q_m\left( Q_n q_n \right) &=& - \left( q_m Q_n \right) q_n +
    q_m ,
\label{sec2-170a}\\
    q_m\left( Q_m q_n \right) &=& - \left( q_m Q_m \right) q_n +
    q_n ,
\label{sec2-170b}\\
    Q_m\left( q_n Q_n \right) &=& - \left( Q_m q_n \right) Q_n +
    Q_m ,
\label{sec2-170c}\\
    Q_m\left( q_m Q_n \right) &=& - \left( Q_m q_m \right) Q_n +
    Q_n .
\label{sec2-170d}
\end{eqnarray}
For the product of three operators as $q(QQ)$ and $Q(qq)$
and with two different indexes $m \neq n$ we have
\begin{eqnarray}
    q_m\left( Q_m Q_n \right) &=& \left( q_m Q_m \right) Q_n +
    \mathcal{H}_{3,1} ,
\label{sec2-170e}\\
    q_m\left( Q_n Q_m \right) &=& \left( q_m Q_n \right) Q_m +
    \mathcal{H}_{3,1}' ,
\label{sec2-170f}\\
    Q_m\left( q_m q_n \right) &=& \left( Q_m q_m \right) q_n +
    \mathcal{H}_{3,2},
\label{sec2-170g}\\
    Q_m\left( q_n q_m \right) &=& \left( Q_m q_n \right) q_m +
    \mathcal{H}_{3,2}' ,
\label{sec2-170h}\\
    Q_m\left( Q_n q_m \right) &=& \left( Q_m Q_n \right) q_m +
    \mathcal{H}_{3,3} ,
\label{sec2-170i}\\
    Q_n\left( Q_m q_m \right) &=& \left( Q_n Q_m \right) q_m +
    \mathcal{H}_{3,3}' ,
\label{sec2-170j}\\
    q_m\left( q_n Q_m \right) &=& \left( q_m q_n \right) Q_m +
    \mathcal{H}_{3,4} ,
\label{sec2-170k}\\
    q_n\left( q_m Q_m \right) &=& \left( q_n q_m \right) Q_m +
    \mathcal{H}_{3,4}' .
\label{sec2-170l}
\end{eqnarray}
The second index $i$ of $\mathcal H_{3,i}$ and $\mathcal H_{3,i}'$ operators means that the corresponding associators (for example, 
$q_m\left( Q_m Q_n \right) - \left( q_m Q_m \right) Q_n = \mathcal{H}_{3,1}$) can be different. We will retain the alternativity property of the octonions
\begin{eqnarray}
    q^2_n Q_m &=& \left( q_n q_n \right) Q_m = q_n \left( q_n Q_m \right),
\label{sec2-180}\\
    q_n \left( Q_m q_n \right) &=& \left( q_n Q_m \right) q_n ,
\label{sec2-190}\\
    Q_m q_n^2 &=& Q_m \left(q_n  q_n \right) = \left(Q_m  q_n \right) q_n .
\label{sec2-200}\\
    Q^2_n q_m &=& \left( Q_n Q_n \right) q_m = Q_n \left( Q_n q_m \right),
\label{sec2-202}\\
    Q_n \left( q_m Q_n \right) &=& \left( Q_n q_m \right) Q_n ,
\label{sec2-204}\\
    q_m Q_n^2 &=& q_m \left(Q_n  Q_n \right) = \left(q_m Q_n \right) Q_n .
\label{sec2-206}
\end{eqnarray}
The associativity rule for the product of four and more operators will be
\begin{equation}
    \hat{a} \left( \hat{b} \hat{c} \right) =
    \pm \left( \hat{a} \hat{b} \right) \hat{c} +
    \mathcal{H}_n \left[ \hat{a} \left( \hat{b} \hat{c} \right) \right]
\label{sec2-210}
\end{equation}
where the operators $\hat{a}, \hat{b}, \hat{c}$ can be any product of
the operators $q_i, Q_j$;
$\mathcal{H}_n \left[ \hat{a} \left( \hat{b} \hat{c} \right) \right]$
is some operator and the sign $(+)$ or $(-)$ in front of the first term of the right hand side of Eq. \eqref{sec2-210} is defined from the multiplication of the corresponding octonions
\begin{equation}
    a \left( bc \right) = \pm \left( ab \right) c
\label{sec2-220}
\end{equation}
where $a,b,c$ are the octonions with the same product of the octonions
$\tilde{q}_i, \tilde{Q}_j$ as in Eq. \eqref{sec2-210}.
\par
One of the main problem in such non-associative algebra with the definitions
\eqref{sec2-72}-\eqref{sec2-206} is the proof of the self-consistency of such algebra.
This is a very complicated problem and we will check the self-consistency for the
products of three operators only.

\vspace{12pt}
\section{The Products of Three Operators}
\vspace{12pt}

At first we will check the non-associativity relationship \eqref{sec2-100}.
For this we will permute the first and third factors in the product
\begin{equation}
\begin{split}
    Q_1 \left( Q_2 Q_3 \right) &= - Q_1 \left( Q_3 Q_2 \right) =
    \left( Q_1 Q_3 \right) Q_2 + \mathcal{H}_3 =
    - \left( Q_3 Q_1 \right) Q_2 + \mathcal{H}_3 \\
    &=
    Q_3 \left( Q_1 Q_2 \right) = - Q_3 \left( Q_2 Q_1 \right) .
\label{sec2-230}
\end{split}
\end{equation}
On the other hand
\begin{equation}
\begin{split}
    Q_1 \left( Q_2 Q_3 \right) &= - \left( Q_1 Q_2 \right) Q_3  + \mathcal{H}_3 =
    \left( Q_2 Q_1 \right) Q_3 + \mathcal{H}_3 =
    - Q_2 \left( Q_1 Q_3 \right) \\
    &=  Q_2 \left( Q_3 Q_1 \right) =
    - \left( Q_2 Q_3 \right) Q_1 + \mathcal{H}_3 =
     \left( Q_3 Q_2 \right) Q_1 + \mathcal{H}_3 \\
     &= -  Q_3 \left(Q_2 Q_1 \right).
\label{sec2-240}
\end{split}
\end{equation}
The same is correct for any permutation of the indexes $(1,2,3)$.
Analogous calculations can be made for the product  with $q_1, q_2, q_3$
\begin{equation}
    q_1 \left( q_2 q_3 \right) =  -  q_3 \left(q_2 q_1 \right).
\label{sec2-250}
\end{equation}
The calculations for three different indexes but with two $q$ and one $Q$
\begin{equation}
    q_1 \left( Q_2 q_3 \right) =  -  q_3 \left(Q_2 q_1 \right).
\label{sec2-250a}
\end{equation}
The same calculations for two $Q$ and one $q$
\begin{equation}
    Q_1 \left( q_2 Q_3 \right) =  -  Q_3 \left(q_2 Q_1 \right).
\label{sec2-250b}
\end{equation}
Eq(s). \eqref{sec2-240}-\eqref{sec2-250b} are correct for any permutation of
the indexes $(1,2,3)$.
The analogous calculations with two equal indexes give us
\begin{eqnarray}
    q_1 \left( Q_1 Q_2 \right) &=& Q_2 \left(Q_1 q_1 \right) - \mathcal{H}_{3,1}' =
    Q_2 \left(Q_1 q_1 \right) + \mathcal{H}_{3,1} - \mathcal{H}_{3,3} - \mathcal{H}_{3,3}',
\label{sec2-260}\\
    Q_1 \left( q_1 Q_2 \right) &=& - Q_2 \left(q_1 Q_1 \right) -
    \mathcal{H}_{3,1} - \mathcal{H}_{3,1}' + i \hbar Q_2 = 
\nonumber \\
    &&- Q_2 \left(q_1 Q_1 \right) -
    \mathcal{H}_{3,3} - \mathcal{H}_{3,3}' + i \hbar Q_2,
\label{sec2-270}\\
    Q_1 \left( q_1 q_2 \right) &=& q_2 \left(q_1 Q_1 \right) - \mathcal{H}_{3,2}' =
    q_2 \left(q_1 Q_1 \right) + \mathcal{H}_{3,2} - \mathcal{H}_{3,4} - \mathcal{H}_{3,4}',
\label{sec2-280}\\
    q_1 \left( Q_1 q_2 \right) &=& - q_2 \left(Q_1 q_1 \right) -
    \mathcal{H}_{3,4} - \mathcal{H}_{3,4}' + i \hbar q_2 = 
\nonumber \\
    &&- q_2 \left(Q_1 q_1 \right) -
    \mathcal{H}_{3,2} - \mathcal{H}_{3,2}' + i \hbar q_2.
\label{sec2-290}
\end{eqnarray}
Consequently on the level of the 3-product one can say only that
\begin{eqnarray}
    \mathcal{H}_{3,1} + \mathcal{H}_{3,1}' &=&
    \mathcal{H}_{3,3} + \mathcal{H}_{3,3}' ,
\label{sec2-300}\\
    \mathcal{H}_{3,2} + \mathcal{H}_{3,2}' &=&
    \mathcal{H}_{3,4} + \mathcal{H}_{3,4}' .
\label{sec2-310}
\end{eqnarray}
The pair of indexes $(1,2)$ in Eq(s). \eqref{sec2-260}-\eqref{sec2-290}
can be replaced by any other pair indexes $(i,j)$. The self-consistency of the relations \eqref{sec2-180}-\eqref{sec2-200} follows from the consistency of the multiplication of the octonions.
\par
On the level of the product of three operators we can not define the operators
$\mathcal{H}_{3,i}, i=1,2,3,4$. But comparing with the relation \eqref{sec2-90} one can
suppose that
\begin{equation}
    \mathcal{H}_{3,i} = \hbar_2 \sum_j
    \left( \alpha_{i,j} q_j + \beta_{i,j} Q_j \right)
\label{sec2-320}
\end{equation}
where $\hbar_2$ is a constant which can be called the second Planck constant
and $\alpha_{i,j} , \beta_{i,j}$ are some constants.

\vspace{24pt}
\section{SOME PROPERTIES OF THE \\ NON-ASSOCIATIVE QUANTUM \\ 
MECHANICS}
\vspace{12pt}

For the octonions $q_i, Q_i$ we have the following relation between $\tilde{q}_i$ and
$\tilde{Q}_i$
\begin{equation}
    \tilde{q}_i \tilde{Q}_i - \tilde{Q}_i \tilde{q}_i= 2 I .
\label{sec3-10}
\end{equation}
This allows us to suppose that the corresponding operators are similar to
canonical conjugate variables and have to have the anti-commutation relationships
\eqref{sec2-90}.
\par
One can see that there is some similarity with between $q_i, Q_j$ and
the annihilation and creation operators of quarks. Every value of index $i=1,2,3$
corresponds to some $color=red, green, blue$. In this case the operators
$q_i, Q_i$ correspond to the annihilation and creation operators of quarks.
But this similarity does not mean the equivalence as the non-associative
quantum mechanics should describe \textit{non-free (strongly interacting) particles}.
\par
In the standard quantum mechanics for the harmonic oscillator one can introduce
the annihilation operator $\hat{a}$ such that
\begin{equation}
    \left. \left. \hat{a} \right| vac_{osc} \right\rangle = 0
\label{sec3-10a}
\end{equation}
where $\left. \left. \right| vac_{osc} \right\rangle$ is a vacuum state for the oscillator.
But for any another potential
\begin{equation}
    \left. \left. \hat{a} \right| vac_{nonosc} \right\rangle \neq 0.
\label{sec3-10b}
\end{equation}
Consequently for the non-associative quantum mechanics we also have
\begin{equation}
    \left. \left. q_i \right| vac_{nonass} \right\rangle \neq 0.
\label{sec3-10c}
\end{equation}
Probably it is similar to the existence of a condensate in quantum chromodynamics. Following the above-mentioned correspondence one can
introduce ``colorless'' operators
$q_1\left( q_2 q_3 \right), Q_1\left( Q_2 Q_3 \right), \left( q_i Q_i \right)$.
These operators can be named by the following manner
\begin{equation}
\begin{split}
     q_1\left( q_2 q_3 \right)  &\text{ -- ``nucleon'' annihilation operator,} \\
     Q_1\left( Q_2 Q_3 \right)  &\text{ -- ``nucleon'' creation operator,} \\
     Q_i q_i                                    &\text{ -- ``meson'' annihilation operator,} \\
     q_i Q_i                                    &\text{ -- ``meson'' creation operator,} \\
     q_i q_i                                    &\text{ -- ``Cooper pair'' annihilation operator,} \\
     Q_i Q_i                                    &\text{ -- ``Cooper pair'' creation operator. }
\label{sec3-20}
\end{split}
\end{equation}

\subsection{The Commutation Relationship for ``meson''}
\vspace{12pt}

Let us consider the commutator
\begin{equation}
    \left( q_i q_i \right) \left( Q_i Q_i \right) -
    \left( Q_i Q_i \right) \left( q_i q_i \right) = ?
\label{sec3-30}
\end{equation}
on $i$ there is no summation. As before we will suppose that the products of four
operators are alternative
\begin{eqnarray}
    q_1 \left( q_1 Q_1^2 \right) &=& q_1^2 Q_1^2 = \left( q_1^2 Q_1 \right) Q_1,
\label{sec3-40}\\
    Q_1 \left( Q_1 q_1^2 \right) &=& Q_1^2 q_1^2 = \left( Q_1^2 q_1 \right) q_1,
\label{sec3-50}\\
    q_1 \left( Q_1^2 q_1 \right) &=& \left( q_1 Q_1^2 \right) q_1.
\label{sec3-60}
\end{eqnarray}
For the another order of the factors
\begin{eqnarray}
    q_1\left( Q_1 \left( q_1 Q_1 \right) \right) &=&
    \left( q_1 Q_1 \right) \left( q_1 Q_1 \right) + \mathcal{H}_{4,1},
\label{sec3-70}\\
    \left( \left( q_1 Q_1 \right) q_1 \right) Q_1 &=&
    \left( q_1 Q_1 \right) \left( q_1 Q_1 \right) + \mathcal{H}_{4,1}',
\label{sec3-80}\\
    Q_1\left( q_1 \left( Q_1 q_1 \right) \right) &=&
    \left( Q_1 q_1 \right) \left( Q_1 q_1 \right) + \mathcal{H}_{4,2},
\label{sec3-90}\\
    \left( \left( Q_1 q_1 \right) Q_1 \right) q_1 &=&
    \left( Q_1 q_1 \right) \left( Q_1 q_1 \right) + \mathcal{H}_{4,2}'.
\label{sec3-100}
\end{eqnarray}
Following to the different ways of the brackets arrangement we have
\begin{equation}
\begin{split}
    q_1^2 Q_1^2 &= Q_1^2 q_1^2 - \mathcal{H}_{4,1} + \mathcal{H}_{4,2} =
    Q_1^2 q_1^2 - \mathcal{H}_{4,1} + \mathcal{H}_{4,2}' \\
    & = Q_1^2 q_1^2 - \mathcal{H}_{4,1}' + \mathcal{H}_{4,2}' =
    Q_1^2 q_1^2 - \mathcal{H}_{4,1}' + \mathcal{H}_{4,2} ,
\label{sec3-110}
\end{split}
\end{equation}
i.e.
\begin{equation}
    \mathcal{H}_{4,1} = \mathcal{H}_{4,1}' , \quad
    \mathcal{H}_{4,2} = \mathcal{H}_{4,2}'
\label{sec3-120}
\end{equation}
and
\begin{equation}
    q_1^2 Q_1^2 - Q_1^2 q_1^2 = - \mathcal{H}_{4,1} + \mathcal{H}_{4,2} .
\label{sec3-130}
\end{equation}
One can suppose that the difference $\mathcal{H}_{4,1} - \mathcal{H}_{4,2}$
can be defined demanding the self-consistency of the product of five operators
and so on.

\vspace{24pt}
\section{CONCLUSIONS AND DISCUSSION}
\vspace{12pt}

In this notice it is shown that one can try to destroy the octonions multiplication
rules to receive a non-associative quantum mechanics which is not equivalent
to the standard quantum mechanics. The distinction is that the algebra
of operators is generated not only by the canonical commutation/anti-commutations  relationships but also by the rules regulating the brackets order in the operators product. The construction of such quantum mechanics is very complicated problem and now we would like
to list some unresolved problems :
\begin{enumerate}
    \item In this notice we show the self-consistency of the product of three operators.
    But the self-consistency of the product of four and more operators is an open
    problem.
    \item Non-associative operators can not have the representation on the functions
    like $\hat{p}_x = -i \hbar \frac{\partial}{\partial x}$ in
    the standard quantum mechanics. In this context the question is how one can
    define eigenstates and eigenfunctions and what are observables of
    the non-associative operators ?
    \item Probably such non-associative quantum mechanics is hardly realized
    in the Nature but one can presuppose that some generalization of the non-associative
    quantum mechanics for the quantum field theory can exist. The matter is that only for 			free fields the
    algebra of quantized fields is known. The algebra of strongly interacting quantum
    fields
    (in some regimes of quantum chromodynamics, quantum gravity and so on) remains unknown. 		It means
    that the algebra of operators of the strongly interacting fields is much more
    complicated the algebra of operators generating only by the canonical 											commutation/anti-commutation 
    relationships and probably the algebra of strongly interacting fields can be
    non-associative one.
    \item One can presuppose that the formulation of the rules for the rearrangement
    of brackets in some non-associative operators product will lead to the appearance
    of a second Planck constant.
\end{enumerate}

\end{document}